\documentclass[11pt,a4paper]{article}
\usepackage[utf8]{inputenc}
\usepackage[T1]{fontenc}
\usepackage{geometry}
\usepackage{amsmath, amssymb}
\usepackage{graphicx}
\graphicspath{{figures/}}
\usepackage{enumitem}
\usepackage{url}
\usepackage[hidelinks]{hyperref}
\usepackage{cite}

\title{Deep Learning-Based Multi-Factor Authentication:\\
A Survey of Biometric and Smart Card Integration Approaches}
\author{Abdelilah Ganmati \and Karim Afdel \and Lahcen Koutti\\
\small Computer Systems \& Vision Laboratory, Faculty of Sciences,\\
\small Ibn Zohr University, Agadir, Morocco\\
\small Emails: a.ganmati@uiz.ac.ma, k.afdel@uiz.ac.ma, l.koutti@uiz.ac.ma}
\date{}
\usepackage{tikz}
\usetikzlibrary{matrix,positioning,calc,arrows.meta,fit}
\usepackage{float} 

\begin{document}
\maketitle

\begin{abstract}
\noindent
In the era of pervasive cyber threats and exponential growth in digital services, the inadequacy of single-factor authentication has become increasingly evident. Multi-Factor Authentication (MFA), which combines knowledge-based factors (passwords, PINs), possession-based factors (smart cards, tokens), and inherence-based factors (biometric traits), has emerged as a robust defense mechanism. Recent breakthroughs in deep learning have transformed the capabilities of biometric systems, enabling higher accuracy, resilience to spoofing, and seamless integration with hardware-based solutions. At the same time, smart card technologies have evolved to include on-chip biometric verification, cryptographic processing, and secure storage, thereby enabling compact and secure multi-factor devices. 

This survey presents a comprehensive synthesis of recent work (2019--2025) at the intersection of deep learning, biometrics, and smart card technologies for MFA. We analyze biometric modalities (face, fingerprint, iris, voice), review hardware-based approaches (smart cards, NFC, TPMs, secure enclaves), and highlight integration strategies for real-world applications such as digital banking, healthcare IoT, and critical infrastructure. Furthermore, we discuss the major challenges that remain open, including usability--security tradeoffs, adversarial attacks on deep learning models, privacy concerns surrounding biometric data, and the need for standardization in MFA deployment. By consolidating current advancements, limitations, and research opportunities, this survey provides a roadmap for designing secure, scalable, and user-friendly authentication frameworks.
\end{abstract}


\section{Introduction and Motivation}

In today’s hyperconnected digital environment, safeguarding user identity has become more critical than ever. With increasing reports of data breaches, phishing scams, and account hijacking, traditional password-based authentication has proven insufficient. Passwords are not only prone to being forgotten or reused, but are also vulnerable to brute-force attacks, phishing, and database leaks. These vulnerabilities have driven the widespread adoption of Multi-Factor Authentication (MFA) as a more robust alternative.

MFA relies on the combination of independent identity factors—something the user knows (e.g., a password), something they have (e.g., a smart card), and something they are (e.g., a biometric trait). This layered approach significantly improves security by ensuring that compromising one factor does not grant access to protected systems. According to recent surveys \cite{TranTruong2025, Mohammed2023}, modern digital services, including financial platforms compliant with regulations such as PSD2, increasingly require MFA to mitigate risks associated with single-factor systems. Privacy-preserving MFA approaches leveraging deep learning and biometrics have also been proposed \cite{Das2020}.

Biometric authentication, including modalities such as face, fingerprint, iris, and voice recognition, has emerged as a popular inherence factor in MFA systems \cite{Jain2011IntroBiometrics}.
 Biometrics offer non-transferable, user-specific traits, improving usability and reducing reliance on memory. Yet, they raise unique challenges in terms of spoofing, privacy, and demographic bias \cite{Ratha2001, Suleski2023}.

Recent advances in deep learning (DL) have significantly improved the accuracy and reliability of biometric systems. Convolutional Neural Networks (CNNs) and other deep architectures have shown strong performance in extracting robust features from noisy or occluded biometric data, enabling real-time and multimodal biometric authentication \cite{WangDeng2021, Zahid2024}. DL techniques also power liveness detection, domain adaptation, and template security, enhancing resilience against spoofing and adversarial attacks \cite{Guo2019}.

In parallel, possession-based factors have evolved. Smart cards, Trusted Platform Modules (TPMs), and secure enclaves now provide cryptographic computation, biometric match-on-card, and tamper-resistant credential storage. Recent advances in biometric payment cards and NFC-enabled devices illustrate how hardware tokens can securely integrate with DL-based biometric authentication to form compact and user-friendly MFA schemes \cite{Patel2019, Tani2025}. These solutions reduce fraud while ensuring compliance with data protection requirements.

Despite this progress, several open challenges remain: ensuring usability without degrading security, defending against presentation attacks, preserving biometric privacy, and ensuring interoperability across vendors and regulatory frameworks \cite{Suleski2023, Mohammed2023}.

\textbf{Objectives.} This survey provides a comprehensive overview of recent research (2019--2025) on DL-based MFA systems integrating biometrics and smart cards. Specifically, it:
\begin{itemize}
    \item Reviews deep learning methods applied to biometric authentication within MFA frameworks.
    \item Examines smart card and hardware-based approaches for secure factor integration.
    \item Compares architectures, fusion strategies, datasets, and benchmarks employed in state-of-the-art systems.
    \item Analyzes threat models and countermeasures against adversarial and spoofing attacks.
    \item Identifies open research questions and outlines future directions for scalable, privacy-preserving, and user-friendly authentication.
\end{itemize}

\section{Background: MFA, Biometrics, and Smart Cards}

Multi-Factor Authentication (MFA) is a security mechanism that requires users to verify their identity using a combination of independent factors, typically categorized as knowledge (something the user knows), possession (something the user has), and inherence (something the user is) \cite{Mohammed2023}. The rationale behind MFA is that compromising multiple independent factors is significantly harder for attackers, thus enhancing overall system security.

Traditionally, authentication relied heavily on passwords. However, due to their susceptibility to guessing, phishing, and large-scale leaks, passwords alone have become increasingly risky. In response, MFA has been adopted across domains handling sensitive data, particularly digital banking, e-government, and healthcare IoT \cite{TranTruong2025, Suleski2023}.

Biometric authentication represents the ``inherence'' factor, leveraging physiological traits (e.g., face, fingerprint, iris) or behavioral patterns (e.g., gait, keystroke dynamics). Unlike passwords or tokens, biometric traits are intrinsic to individuals and difficult to replicate. However, they raise challenges such as privacy concerns, irrevocability, and vulnerability to spoofing attacks \cite{Ratha2001, Zahid2024}. To mitigate these risks, standardization bodies (e.g., ISO/IEC 30107) have proposed presentation attack detection (PAD) guidelines, and researchers are increasingly focused on fairness and robustness across demographics \cite{Alduhailan2025}.

Smart cards, typically associated with the ``possession'' factor, are tamper-resistant hardware devices capable of securely storing credentials, cryptographic keys, and even performing biometric matching. In the banking sector, EMV-compliant cards enable secure offline authentication through digital signatures. When combined with biometrics, smart cards can implement match-on-card verification, ensuring that sensitive templates never leave the card’s secure chip \cite{Tani2025, Patel2019}.

Beyond traditional smart cards, Trusted Platform Modules (TPMs) and Secure Enclaves extend these guarantees to general-purpose devices such as smartphones and laptops. These components isolate sensitive data and computations, enabling secure biometric enrollment, inference, and key storage, and underpin modern standards such as FIDO2 and WebAuthn.

In practice, effective MFA requires balancing usability, cost, and risk. A typical modern system may combine: a fingerprint scan (inherence), a smartphone secure enclave or biometric smart card (possession), and a PIN or behavioral pattern (knowledge/behavior). The integration of deep learning into biometric systems---coupled with trusted hardware---marks the next evolution in MFA, explored in detail in the following sections.

\section{Deep Learning for Biometric Authentication}

Deep learning (DL) has fundamentally transformed biometric authentication by enabling end-to-end learning, robust feature extraction, and scalability across diverse modalities. Traditional biometric systems relied on handcrafted features, which often lacked generalizability across populations or environmental conditions. DL models --- particularly Convolutional Neural Networks (CNNs), Recurrent Neural Networks (RNNs), and Transformers --- now power state-of-the-art systems for face, fingerprint, iris, voice, and behavioral biometrics \cite{Alduhailan2025,WangDeng2021,Zahid2024}.

\subsection{Facial Recognition and Anti-Spoofing}
Facial recognition has rapidly advanced with architectures such as FaceNet \cite{Schroff2015}, ArcFace \cite{Deng2019}, and CosFace \cite{Wang2018}, which learn highly discriminative embeddings from images. Wang and Deng \cite{WangDeng2021} survey modern DL-based face recognition.
  
However, face systems remain vulnerable to spoofing (2D photos, replay videos, 3D masks). Liveness detection networks address this by analyzing texture, motion cues, or physiological signals (e.g., eye blinking, rPPG). Guo et al. \cite{Guo2019} proposed CNN-based liveness detection, while recent works integrate anti-deepfake detection.

\begin{figure}[h]
\centering
\includegraphics[width=0.85\linewidth]{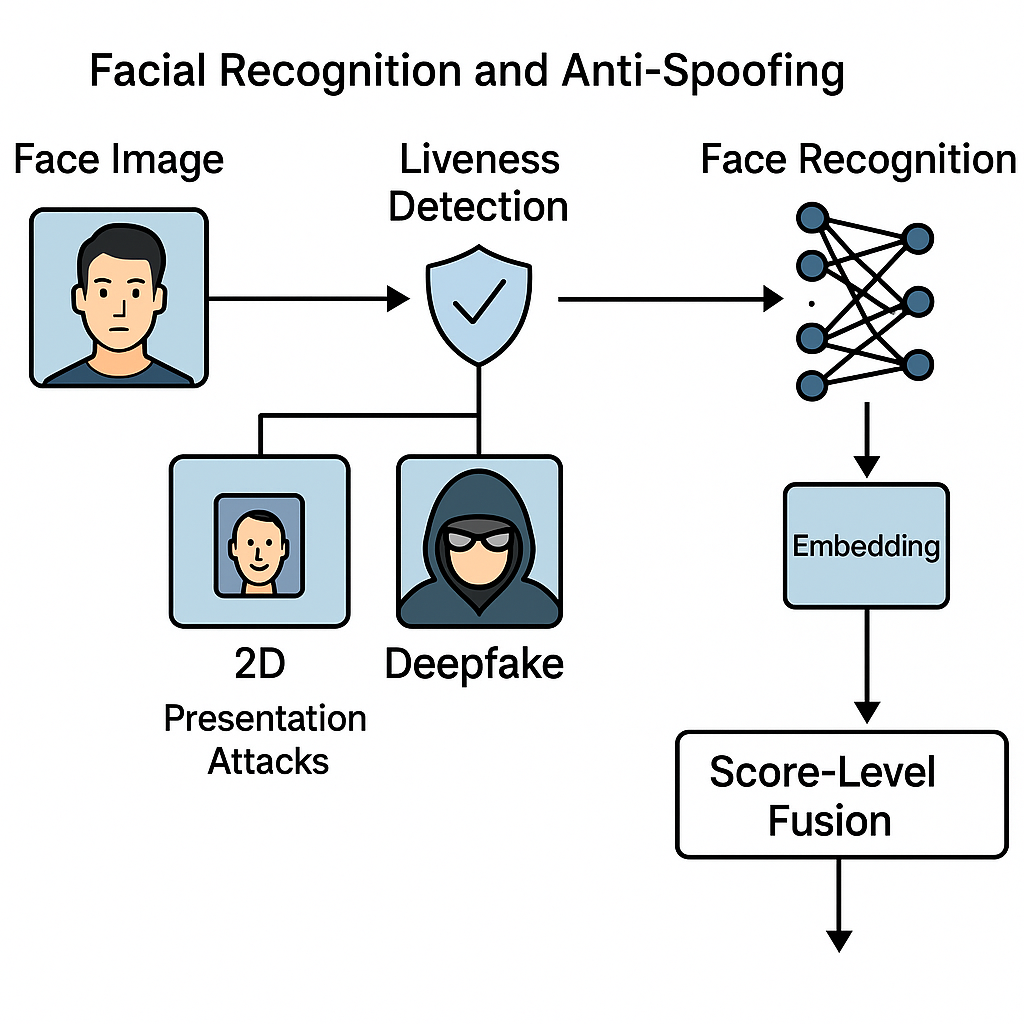}
\caption{Multimodal biometric fusion with score-level fusion. Columns are fixed on a grid; all connectors use right-angle routing to avoid overlap.}
\label{fig:multimodal_spoofing fusion}
\end{figure}

\subsection{Fingerprint and Iris Recognition}
DL enhances fingerprint authentication by improving minutiae detection, ridge classification, and partial print matching. Zahid et al. \cite{Zahid2024} show CNNs outperform traditional Gabor-based methods under noisy conditions.  
Similarly, iris recognition benefits from CNNs and attention-based models trained on datasets such as CASIA-Iris and ND-Iris, robust to illumination and pupil dilation \cite{Qureshi2024}.

\subsection{Voice and Behavioral Biometrics}
DL also advances speaker verification via spectrogram-based CNNs and LSTM embeddings. Combining voice with face (audiovisual biometrics) strengthens robustness for remote banking authentication \cite{Vatchala2024}.  
Behavioral biometrics --- keystroke, gait, touchscreen, mouse movement --- enable continuous MFA. Verma et al. \cite{Verma2022} demonstrate smartphone motion-based DL models for adaptive MFA.

\subsection{Multimodal and Fusion Architectures}
Fusing modalities (e.g., face + voice, fingerprint + iris) improves both accuracy and spoofing resistance. DL supports early (input-level), intermediate (embedding-level), and late (score-level) fusion \cite{Damer2019}.  
Talreja et al. \cite{Talreja2019} proposed deep multibiometric hashing to unify modalities into compact binary templates for efficiency.

\begin{figure}[h]
\centering
\includegraphics[width=0.85\linewidth]{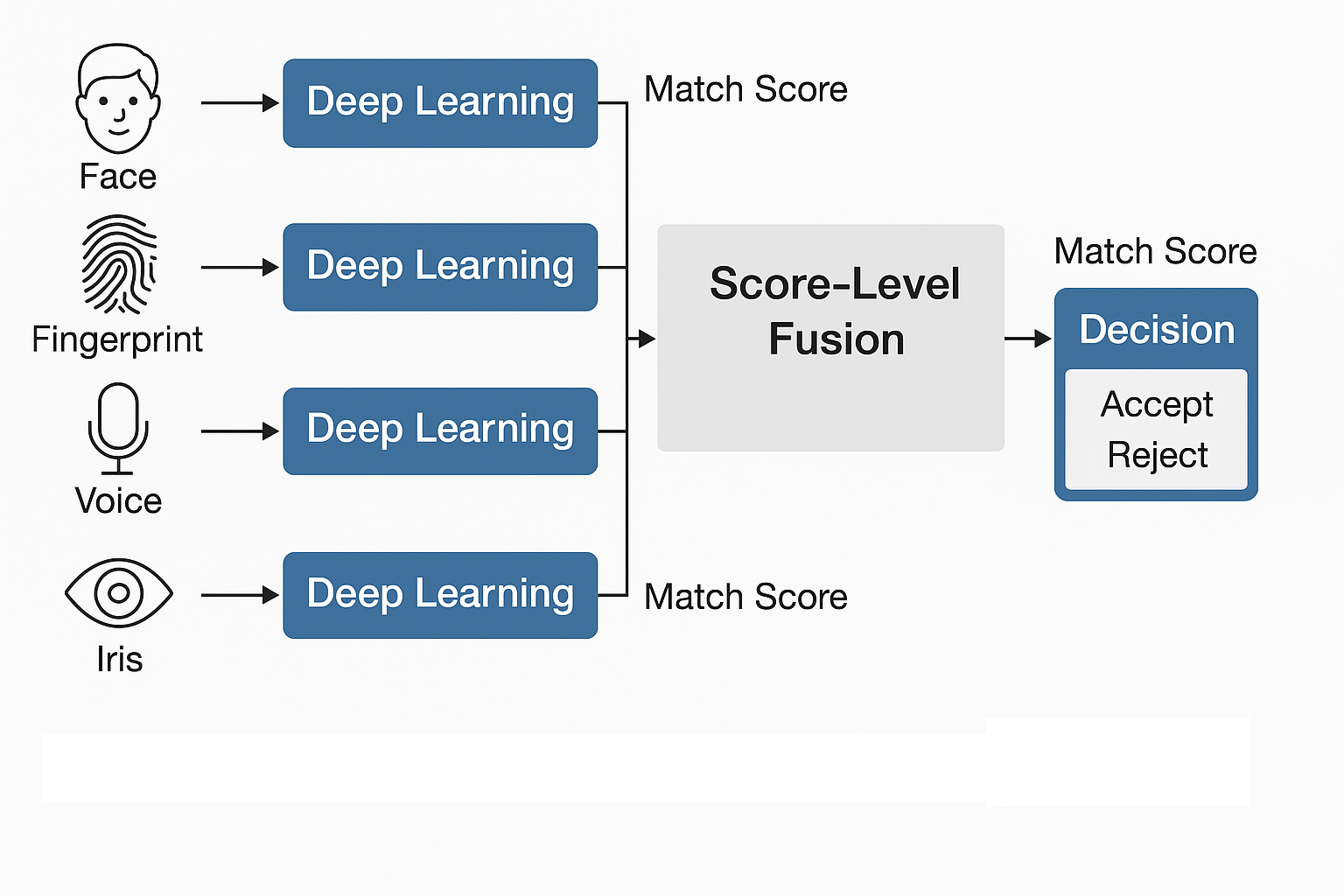}
\caption{Multimodal biometric fusion architecture using DL.}
\label{fig:multimodal_fusion}
\end{figure}

\subsection{Liveness Detection and Anti-Spoofing}
Anti-spoofing is now integral to MFA. Approaches include texture-based CNNs, challenge-response (blink, lip movement), and rPPG estimation. Lee et al. \cite{Lee2023} propose integrating DL-based anti-spoofing directly in biometric smart cards.

\subsection{Trends and Challenges}
Despite advances, challenges remain:  
\begin{itemize}
    \item Dataset bias and fairness across demographics.  
    \item Adversarial attacks against DL models.  
    \item Lightweight deployment on constrained hardware (e.g., smart cards, IoT).  
    \item Lack of explainability and interpretability.  
\end{itemize}
Nonetheless, DL has enabled robust, multimodal, and real-time biometric MFA systems.

\begin{table}[h]
\centering
\caption{Representative DL models across biometric modalities}
\label{tab:dl_modalities}
\begin{tabular}{|p{2.5cm}|p{4.5cm}|p{3.5cm}|}
\hline
\textbf{Modality} & \textbf{DL Models / Techniques} & \textbf{Key References} \\
\hline
Face & ArcFace, CosFace, FaceNet; CNNs for liveness & \cite{WangDeng2021,Guo2019} \\
\hline
Fingerprint & CNN minutiae detection, ridge classifiers & \cite{Zahid2024} \\
\hline
Iris & CNN/attention on CASIA, ND-Iris datasets & \cite{Qureshi2024} \\
\hline
Voice & Spectrogram CNNs, LSTM embeddings & \cite{Vatchala2024} \\
\hline
Behavioral & Keystroke, gait, smartphone motion via DL & \cite{Verma2022} \\
\hline
Multimodal & Embedding/score fusion, multibiometric hashing & \cite{Damer2019,Talreja2019} \\
\hline
\end{tabular}
\end{table}

\section{Hardware Factors and Smart Card Integration}

Biometric authentication becomes significantly more robust when coupled with hardware-based security mechanisms such as smart cards, Trusted Execution Environments (TEEs), and Trusted Platform Modules (TPMs). These components not only enable secure storage of biometric templates and cryptographic keys but also support tamper-resistant execution of verification algorithms. The synergy between biometric recognition and possession-based hardware is foundational to secure MFA.

\subsection{Smart Cards in MFA Systems}

Smart cards have long been used in authentication due to their ability to securely store user credentials and perform local computations. In modern MFA, biometric smart cards (BSCs) integrate fingerprint or facial recognition sensors directly into the card or terminal.

There are two main architectures:
\begin{itemize}
    \item \textbf{Match-on-Card (MoC):} Biometric matching is performed entirely on the card's chip, and the template never leaves the card. This ensures maximum privacy.
    \item \textbf{Match-off-Card:} The biometric is matched externally, with the template read from the card. This mode is more flexible but less private.
\end{itemize}

MoC provides superior privacy and is increasingly supported by commercial products such as Idemia’s biometric payment cards and Gemalto’s biometric EMV solutions, with technical standards maintained by EMVCo \cite{Patel2019,Sabri2019,EMVCo2021}.

\begin{table}[h]
\centering
\caption{Smart Card Types and Capabilities in Biometric MFA}
\label{tab:smartcards}
\begin{tabular}{|p{3cm}|p{3cm}|p{2.5cm}|p{3cm}|p{3cm}|}
\hline
\textbf{Type} & \textbf{Biometric Support} & \textbf{Match Mode} & \textbf{Cryptographic Support} & \textbf{Example Use Case} \\
\hline
Basic Smart Card & None & -- & Symmetric/Asymmetric keys & OTP authentication, access control \\
\hline
Biometric Smart Card & Fingerprint/Face & Match-on-Card & ECC/RSA, Secure Enclave & e-ID, biometric payment cards \\
\hline
TPM-Backed Card & External biometric reader & Match-off-Card & TPM, Secure Boot & Government ID, enterprise logins \\
\hline
Hybrid MFA Card & Face + PIN/Fingerprint & Dual Match & Secure Element + DL & High-security financial systems \\
\hline
\end{tabular}
\end{table}

\subsection{Secure Enclaves and Trusted Execution}

Modern smartphones, tablets, and IoT devices increasingly include TEEs (e.g., ARM TrustZone, Intel SGX) or TPMs (Trusted Platform Modules), which isolate cryptographic operations and template storage from the main OS. TPMs ensure that private keys and biometric hashes are never exposed to system memory. TEEs can even run DL inference models securely, enabling on-device verification. This convergence of DL inference and hardware-backed isolation enables privacy-preserving authentication even on untrusted networks \cite{Luo2020,Suleski2023}.

\subsection{FIDO2 and WebAuthn Protocols}

Standardization is crucial for interoperability. The FIDO2 framework, including WebAuthn and CTAP, allows biometric authentication via trusted devices without transmitting secrets to remote servers. In this model:
\begin{enumerate}
    \item A biometric scan (e.g., fingerprint) unlocks a private key stored in the authenticator (smart card or phone).
    \item The key signs a challenge and proves identity to the relying party.
\end{enumerate}

FIDO2-compliant authenticators such as YubiKey Bio and Windows Hello incorporate DL-based biometric recognition, aligning with privacy-by-design principles \cite{FIDO2022,Mohammed2023}.

\subsection{Smart Card + DL System Architectures}

Recent architectures combine deep learning with secure hardware to enhance biometric verification:
\begin{itemize}
    \item \textbf{DL on-chip:} Miniaturized CNNs embedded in smart cards or tokens.
    \item \textbf{Secure template fusion:} Combining multiple traits (e.g., fingerprint + iris) with fused templates stored in secure elements.
    \item \textbf{On-device adaptation:} DL models fine-tuned per user during enrollment, stored within TEEs.
\end{itemize}

Tani et al. (2025) validated the feasibility of such architectures for real-world banking authentication \cite{Tani2025}.

\subsection{Challenges in Hardware-Based MFA}

Despite their advantages, hardware-integrated MFA systems face challenges:
\begin{itemize}
    \item \textbf{Cost and Scalability:} Biometric smart cards are more expensive than traditional tokens.
    \item \textbf{Hardware Standardization:} Fragmentation across vendors complicates integration.
    \item \textbf{Energy Constraints:} DL inference on low-power chips requires lightweight models and quantization.
\end{itemize}

Ongoing research explores energy-efficient DL models and secure co-processors to address these limitations \cite{Lee2023,AlAssam2010}.

\section{System Integration Patterns}

The integration of deep learning (DL)--based biometric authentication with traditional MFA systems introduces several design options depending on modality fusion, user experience (UX) requirements, and environmental constraints. This section discusses core system-level integration strategies, such as fusion techniques, adaptive MFA policies, and UX considerations like latency.

\subsection{Fusion Techniques in DL-Based MFA}

To leverage the strengths of multiple authentication factors---especially in multimodal biometrics---systems often use fusion strategies to combine different sources of evidence. Fusion can occur at various levels \cite{Damer2019}:

\begin{itemize}
    \item \textbf{Sensor-Level Fusion:} Raw biometric signals (e.g., fingerprint + face) are captured and preprocessed jointly.
    \item \textbf{Feature-Level Fusion:} Deep embeddings from CNN/RNN models are concatenated before classification.
    \item \textbf{Score-Level Fusion:} Independent DL models output match scores that are weighted and combined.
    \item \textbf{Decision-Level Fusion:} Each modality votes independently; a decision is made based on predefined logic (e.g., majority voting).
\end{itemize}

Score-level fusion offers a balance between flexibility and performance, and is widely used in commercial systems. Decision-level fusion is preferred in scenarios with hardware heterogeneity or legacy compatibility.

\begin{table}[h]
\centering
\caption{Fusion strategies in DL-based MFA systems}
\label{tab:fusion_strategies}
\begin{tabular}{|p{3cm}|p{3cm}|p{4cm}|p{4cm}|}
\hline
\textbf{Fusion Level} & \textbf{DL Involvement} & \textbf{Pros} & \textbf{Challenges / Example Use Case} \\
\hline
Sensor-Level & Preprocessing with CNNs & Rich feature interaction & Hardware synchronization; face+voice capture in mobile apps \\
\hline
Feature-Level & Embedding concatenation & High discriminative power & Risk of overfitting; face+iris CNN embeddings \\
\hline
Score-Level & Score normalization + DL & Modularity, scalable & Needs calibration; face+fingerprint fusion in banking apps \\
\hline
Decision-Level & Independent DL classifiers & Robust to missing modalities & May lose subtle correlations; face+PIN at ATMs \\
\hline
\end{tabular}
\end{table}

\subsection{Risk-Adaptive and Contextual MFA}

Static MFA policies (e.g., always requiring password + fingerprint) can frustrate users in low-risk contexts and may still be insufficient in high-risk situations. Risk-adaptive MFA (RA-MFA) tailors factor requirements based on contextual and behavioral signals \cite{Verma2022,Suleski2023}.

Key features include:
\begin{itemize}
    \item \textbf{Context-aware triggers:} Location, device, IP reputation, login time.
    \item \textbf{Behavioral biometrics:} Gait, touch dynamics, keystroke rhythm monitored in background.
    \item \textbf{Continuous authentication:} Ongoing risk scoring (e.g., via RNNs) during a session.
\end{itemize}

Risk-based MFA reduces friction by escalating only when needed. For example, a face + PIN may suffice from a known device, but access from a new region might prompt for a smart card tap or one-time password (OTP). DL models are increasingly used to learn context-aware risk patterns and behavioral baselines \cite{Gordon2018}.

\begin{figure}[h]
\centering
\includegraphics[width=0.8\linewidth]{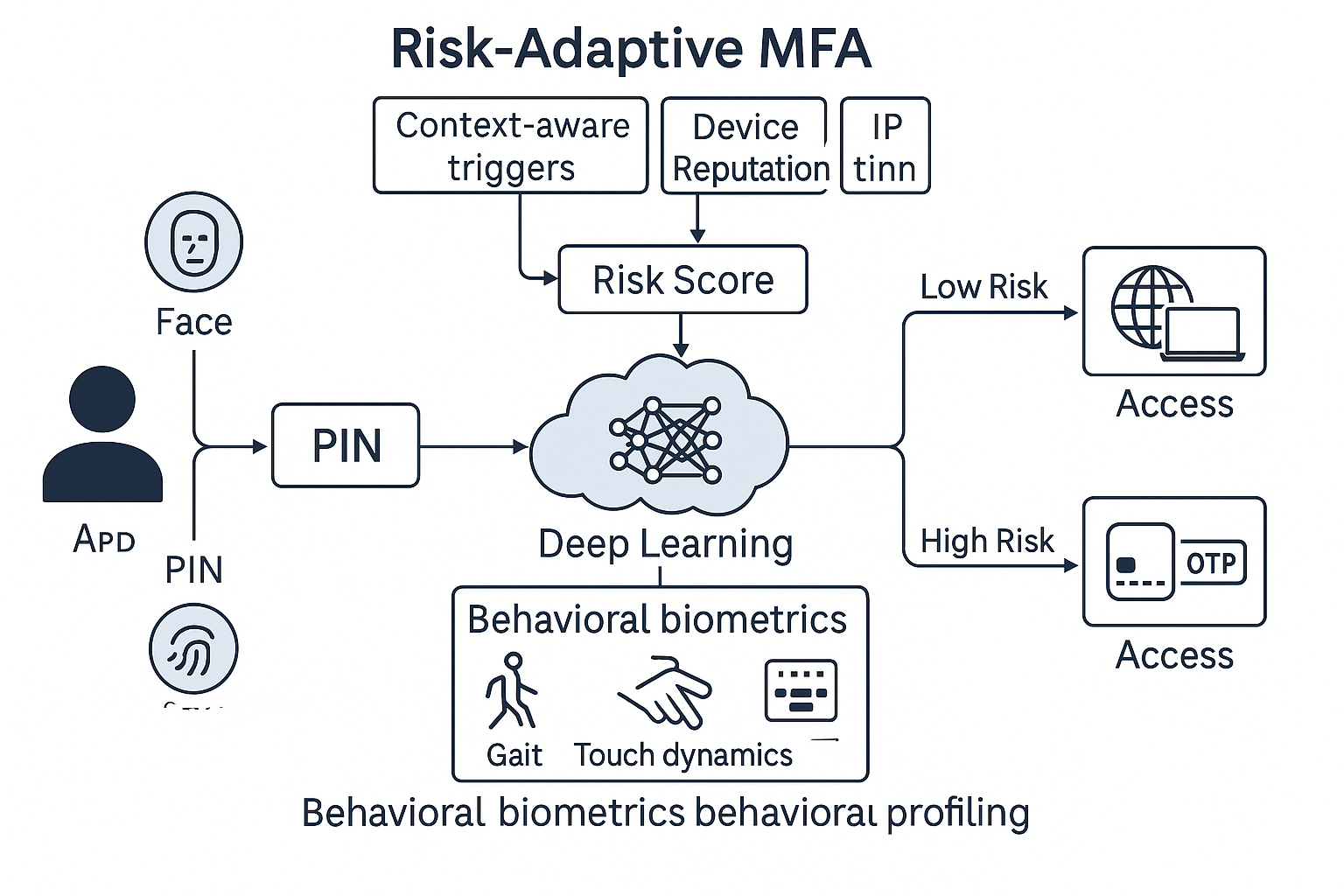}
\caption{Risk-adaptive MFA flow integrating deep learning and behavioral profiling.}
\label{fig:risk_adaptive}
\end{figure}

\subsection{Latency, UX, and Scalability Considerations}

Real-world deployments must balance security, usability, and performance. High authentication accuracy is irrelevant if latency leads to user abandonment.

Important UX metrics include:
\begin{itemize}
    \item \textbf{Time-to-Authenticate:} Ideally $<1.5$ seconds for biometric verification.
    \item \textbf{False Reject Rate (FRR):} Should not exceed 2--3\% for frequent logins.
    \item \textbf{Fallback Experience:} Password reset or re-authentication must not nullify MFA strength.
\end{itemize}

Strategies to improve UX:
\begin{itemize}
    \item On-device DL inference (MobileNet, EfficientNet-Lite).
    \item Asynchronous verification for non-critical actions.
    \item Secure enclave caching to avoid repeated scans.
\end{itemize}

Edge-optimized DL frameworks (TensorFlow Lite, CoreML, PyTorch Mobile) enable smooth deployment in mobile and smartcard MFA \cite{Tani2025}.

\section{Open Challenges and Threat Models}

While DL-based MFA systems offer enhanced security and usability, their integration introduces new vulnerabilities, operational hurdles, and ethical concerns. This section explores key challenges validated in recent literature and standardized frameworks.

\subsection{Adversarial Robustness and Deepfake Spoofing}

Deep learning models, particularly those used in facial and voice biometrics, are susceptible to adversarial manipulation—small perturbations or AI-generated forgeries that can bypass recognition systems.  
Goodfellow et al. \cite{Goodfellow2015} formally described adversarial examples in neural networks, while more recent surveys show how these attacks threaten biometric verification systems \cite{Akhtar2018, Deb2020}.

\textbf{Adversarial Attacks:}  
Subtle perturbations in input images can cause CNNs to misclassify identities. Physical-world attacks (e.g., adversarial glasses, printed patches) have been experimentally shown to deceive face recognition systems.

\textbf{Deepfakes:}  
AI-based synthesis of faces or voices can impersonate real users. As noted by Deb et al. \cite{Deb2020}, deepfake attacks bypass liveness checks unless rPPG or challenge-response features are integrated.

\textbf{Presentation Attacks (PAs):}  
Traditional 2D photo, mask, and replay attacks remain among the most common spoofing methods. The ISO/IEC 30107 standard formalizes PAD (Presentation Attack Detection) testing protocols \cite{ISO30107}.

\textbf{Mitigation:}  
Robust countermeasures include adversarial training, attention-guided liveness detection, and cross-modal consistency (e.g., face + voice validation). Hybrid anti-spoofing solutions combining texture and temporal cues show promising defense potential \cite{Qureshi2024, Lee2023}.

\begin{table}[H]
\centering
\caption{Verified threats and mitigation strategies in DL-based MFA systems}
\label{tab:threats}
\begin{tabular}{|p{3cm}|p{4cm}|p{3cm}|p{5cm}|}
\hline
\textbf{Threat Type} & \textbf{Example Attack} & \textbf{Vulnerable Modality} & \textbf{Mitigation Strategy / Source} \\
\hline
Adversarial Input & Pixel perturbations, adversarial patches & Face, Iris & Adversarial training, gradient masking \cite{Goodfellow2015, Akhtar2018} \\
\hline
Deepfake Spoofing & Synthetic faces or voices & Face, Voice & DL-based liveness, multimodal consistency \cite{Deb2020, Qureshi2024} \\
\hline
Presentation Attack & Printed photo, silicone mask & Face, Fingerprint & PAD-compliant testing (ISO/IEC 30107) \cite{ISO30107} \\
\hline
Replay Attack & Recorded voice/video & Voice, Gait & Challenge-response protocols, audio watermarking \cite{Verma2022} \\
\hline
Sensor-Level Injection & Tampered biometric input & All & Trusted sensors, hardware-based validation \cite{Luo2020, Tani2025} \\
\hline
\end{tabular}
\end{table}

\subsection{Privacy and Biometric Template Protection}

Unlike passwords, biometric data is irrevocable—once compromised, it cannot be reset.  
Recent research highlights this as a central ethical issue for DL-based MFA systems.  
Key privacy-preserving mechanisms include:

\begin{itemize}
    \item \textbf{Cancelable Biometrics:} Transformations that make stored templates non-invertible, as defined by Ratha et al. \cite{Ratha2001}.
    \item \textbf{Homomorphic Encryption:} Enables matching of encrypted features without exposing raw biometric data \cite{Das2020}.
    \item \textbf{Secure Enclaves and Match-on-Card:} Local verification using TPMs or smart cards ensures templates never leave trusted hardware \cite{Tani2025, FIDO2022}.
\end{itemize}

These approaches align with GDPR and NIST SP 800-63B guidelines for identity management and privacy-by-design \cite{NIST80063B}.

\subsection{Bias, Fairness, and Inclusivity}

Deep learning models can encode biases due to unbalanced training datasets.  
Empirical results from Buolamwini and Gebru \cite{Buolamwini2018} demonstrated that commercial facial recognition systems exhibit higher false rejection rates for darker-skinned females.  
This has since prompted fairness auditing and standardized demographic testing (ISO/IEC TR 22116).  
Mitigation strategies include bias-aware loss functions, adversarial debiasing, and diverse data augmentation \cite{Mohammed2023}.

\subsection{Standardization and Interoperability}

DL-MFA systems depend on multiple heterogeneous components—sensors, models, hardware, and communication protocols.  
To ensure interoperability, systems must adhere to recognized standards:

\begin{itemize}
    \item \textbf{FIDO2/WebAuthn:} Enables passwordless authentication with biometrics and public-key cryptography \cite{FIDO2022}.
    \item \textbf{ISO/IEC 30107:} Defines PAD (Presentation Attack Detection) and evaluation criteria for anti-spoofing.
    \item \textbf{ISO/IEC 19795:} Governs performance testing and reporting in biometric systems.
\end{itemize}

Standard compliance enhances cross-vendor compatibility and ensures reproducible evaluation results.

\subsection{Summary}

DL-based MFA improves identity assurance but still faces challenges in adversarial robustness, deepfake resistance, privacy, and bias mitigation.  
The future of trustworthy MFA depends on standardized testing, open datasets, and secure hardware-software co-design.

\subsection{Standardization and Interoperability}

DL-MFA systems often combine heterogeneous sensors, inference engines, and secure hardware. Without standards, integration is fragile and error-prone.  
Key standards include: FIDO2/WebAuthn, ISO/IEC 30107 (PAD), and EMVCo for biometric payment cards \cite{FIDO2022,Suleski2023}.  
Future work must emphasize pluggable frameworks, formal verification of workflows, and government-led certification (e.g., eIDAS, NIST 800-63).  

\subsection{Summary}

While DL-based MFA significantly strengthens digital identity protection, its adoption introduces challenges in robustness, privacy, fairness, and interoperability. Addressing these threats requires cross-disciplinary efforts spanning ML research, hardware security, regulatory compliance, and usability studies.

\section{Datasets, Benchmarks, and Metrics}

Robust evaluation of DL-based MFA systems requires standardized biometric datasets and consistent performance metrics. This section presents widely used benchmarks for facial, fingerprint, iris, and multimodal biometrics, and outlines key evaluation criteria such as FAR, FRR, and EER.

\subsection{Benchmark Datasets for Biometric MFA}

Research in DL-powered biometric authentication relies on curated datasets that represent different modalities under diverse conditions. The quality and bias of these datasets significantly influence model generalizability.

Well-established datasets underpin biometric research. Examples include LFW \cite{LFW2007}, VGGFace2 \cite{Cao2018}, AgeDB \cite{Moschoglou2017}, CASIA-IrisV4 \cite{CASIA2010}, and FVC2004 \cite{FVC2004}, which have become de facto benchmarks for DL-based evaluation.

\begin{table}[h]
\centering
\caption{Commonly Used Datasets for DL-Based MFA Evaluation}
\label{tab:datasets}
\begin{tabular}{|p{2.8cm}|p{2cm}|p{2cm}|p{4cm}|p{4cm}|}
\hline
\textbf{Dataset} & \textbf{Modality} & \textbf{Size} & \textbf{Features} & \textbf{Use Case} \\
\hline
LFW \cite{LFW2007} & Face & 13K images & Unconstrained faces & Face verification, CNN evaluation \\
\hline
VGGFace2 \cite{VGGFace22018} & Face & 3.3M images & High intra-class variability & Deep embedding learning, pose/age invariance \\
\hline
AgeDB \cite{AgeDB2017} & Face & 12K images & Age progression benchmark & Age-invariant face verification \\
\hline
CASIA-IrisV4 \cite{CASIAIrisV4} & Iris & 54K samples & NIR iris images, multiple views & Iris recognition \\
\hline
FVC2004 \cite{FVC2004} & Fingerprint & 3.6K images & Noisy, rotated prints & Fingerprint verification, liveness detection \\
\hline
PLUSVein-FV3 \cite{PlusVein2021} & Finger Vein & 2K+ samples & Dorsal hand vein patterns & Vein-based biometric verification \\
\hline
BIOMDATA \cite{BIOMDATA2015} & Multimodal & 1.2K samples & Iris, fingerprint, face & Fusion-based authentication \\
\hline
SOTAMD \cite{SOTAMD2022} & Spoofing & $\sim$5K samples & Presentation attacks & Anti-spoof model training \\
\hline
\end{tabular}
\end{table}

Many of these datasets include variations in lighting, pose, aging, and acquisition devices to simulate real-world conditions. Public availability supports benchmarking and fair comparison across DL architectures.

\subsection{Performance Metrics in MFA Systems}

Accurate evaluation of DL-based MFA systems requires standardized biometric metrics and protocols, as defined in ISO/IEC 19795 \cite{ISO19795} and NIST SP 800-63B \cite{NIST80063B}.  
These standards ensure comparability across algorithms, datasets, and deployment environments.

\begin{itemize}
    \item \textbf{False Acceptance Rate (FAR):} Probability that an impostor is incorrectly accepted as a genuine user. FAR is critical for measuring system security and is often reported at operating points such as FAR = $10^{-3}$ or $10^{-4}$.
    \item \textbf{False Rejection Rate (FRR):} Probability that a legitimate user is incorrectly rejected. FRR reflects system usability and user experience.
    \item \textbf{Equal Error Rate (EER):} The rate at which FAR and FRR are equal, often used as a single scalar indicator of performance. Lower EER implies a more accurate system.
    \item \textbf{Receiver Operating Characteristic (ROC):} A curve plotting the trade-off between FAR and True Positive Rate (1–FRR). Widely used to visualize model discriminability.
    \item \textbf{Detection Error Tradeoff (DET):} A log-scaled version of the ROC curve emphasizing low-error regions, recommended by ISO/IEC 19795 for biometric evaluations.
    \item \textbf{Failure to Enroll (FTE) / Failure to Acquire (FTA):} Rates at which biometric data cannot be captured or enrolled successfully, critical for deployment evaluation.
    \item \textbf{Authentication Latency:} The time required to complete an MFA process (biometric + token verification). Recommended latency for practical systems is below 1.5 seconds.
\end{itemize}

In multimodal DL-MFA, researchers also evaluate:
\begin{itemize}
    \item \textbf{Fusion Gain:} Improvement in EER or accuracy when combining multiple modalities (e.g., face + fingerprint) compared to single-modality baselines.
    \item \textbf{Robustness to Noise and Spoofing:} Performance degradation under environmental noise, aging, or adversarial conditions \cite{Qureshi2024}.
    \item \textbf{Template Security Impact:} Trade-offs between encryption, privacy-preserving operations, and system accuracy.
\end{itemize}

To ensure reproducibility, public benchmarks such as FVC2004, CASIA-IrisV4, and AgeDB are typically evaluated using these metrics under standard protocols \cite{FVC2004, CASIAIrisV4, AgeDB2017}.

\section{Conclusion and Outlook}

As digital ecosystems expand and cyber threats evolve, MFA has become a central pillar of secure access control. This survey reviewed how DL, biometric modalities, and hardware tokens (e.g., smart cards, secure enclaves) can converge to build the next generation of MFA systems. We highlighted improvements in accuracy, liveness detection, and multimodal fusion; and we identified challenges around robustness, privacy, fairness, and interoperability.

\subsection*{Future Research Roadmap}

\begin{table}[h]
\centering
\caption{Roadmap for Future DL-Based MFA Research}
\label{tab:roadmap}
\begin{tabular}{|p{2.5cm}|p{3.2cm}|p{3.2cm}|p{3.2cm}|p{3.5cm}|}
\hline
\textbf{Domain} & \textbf{Challenge} & \textbf{Role of Deep Learning} & \textbf{Role of Hardware} & \textbf{Future Direction} \\
\hline
Edge Computing & Latency and resource limits & Lightweight CNNs, quantization, TinyML & Smartcards with embedded accelerators & Real-time on-device biometric MFA \\
\hline
Adaptive MFA & Context-aware risk & Behavioral biometrics, RNNs for profiling & Secure enclaves for storage & Risk-based, continuous MFA systems \\
\hline
Privacy & Template leakage & Cancelable biometrics, homomorphic inference & Match-on-card, TPM protection & Privacy-preserving, GDPR-compliant MFA \\
\hline
Security & Adversarial/deepfake spoofing & Robust training, liveness via DL, anomaly detection & Trusted sensors, secure co-processors & Resilient MFA against AI-driven attacks \\
\hline
Trust & Fairness and explainability & Bias auditing, explainable AI & Standardized hardware protocols & Inclusive, transparent MFA adoption \\
\hline
\end{tabular}
\end{table}

\paragraph{Acknowledgments}
This work was conducted at the Computer Systems \& Vision Laboratory, Faculty of Sciences, Ibn Zohr University, Agadir, Morocco.

\bibliographystyle{IEEEtran}
\bibliography{refs}

\end{document}